\documentclass[a4paper,noshowpacs,showkeys,preprint,superscriptaddress,aps,prb,longbibliography]{revtex4-1}
\pdfoutput=1
\usepackage{graphicx}
\usepackage{amssymb}
\usepackage{amsmath}
\usepackage{epsfig}

\usepackage[usenames]{color}

\usepackage[top=1.5in, bottom=1.25in, left=1in, right=1in]{geometry}

\begin{document}

\title{Tailoring one-dimensional layered metamaterials to achieve unidirectional transmission and reflection}

\author{D.~Psiachos}
\email{dpsiachos@gmail.com}
\author{M.~M.~Sigalas}
\email{sigalas@upatras.gr}
\affiliation{Department of Materials Science, University of Patras, 26504, Rio, Greece}

\keywords{elastic wave propagation, metamaterials}
\begin{abstract}
We investigate elastic-wave propagation in a spatially-dispersive multilayered, totally passive
metamaterial system. At oblique incidence a longitudinal (acoustic) wave can convert to transverse in the solid material comprising
the layers, but when the incident wave enters the multilayer from a solid as opposed to a liquid medium, 
the incident transverse component supported by the solid medium
indirectly causes the longitudinal transmission response to be greatly modified and similarly for the
transverse wave exiting the multilayer into a solid medium in response to an incident longitudinal 
 wave. The conversion between longitudinal and transverse waves is found to lead to the emulation of 
a characteristic non-reciprocal phenomenon at some frequencies: a directionality in the
transmission response, sometimes simultaneously with the reflection response. The directionality
can be exploited for example in the construction of antiseismic structures or breakwater structures. The
inclusion of gain/loss elements can strongly enhance the directionality. Periodicity-breaking
defects can cause a great variability in the response, enabling the use of devices based on this phenomenon as sensors.

\end{abstract}

\maketitle

\section{Introduction}
Controlling elastic wave propagation in layered media has been the scope of pioneering theoretical
and recently, experimental studies. While the propagation of acoustic waves
in phononic lattices has many parallels with the propagation
of electromagnetic waves in photonic lattices, showing similar
phenomena and applications, such as acoustic cloaking~\cite{ChenChanAppPh} or illusions~\cite{Layman}, or more generally,
transformation acoustics~\cite{MiltonNJP,NorrisReview}, the more general case of wave propagation in elastodynamic media
is, as yet, not widely studied. 

The interconversion between longitudinal and transverse 
modes has been exploited in the design of metasurfaces in order to emulate nonreciprocal phenomena
in electromagnetic systems~\cite{tangential1,tangential2}.

Non-reciprocal effects such as asymmetric transmission may be emulated by spatially-dispersive~\cite{Helbig} 
metasurfaces. In Ref.~\onlinecite{tangential1} metasurfaces comprised of asymmetrically-aligned 
electric dipoles were studied analytically and numerically with respect to their effect
on the polarization of the incident electromagnetic radiation. It was found that for some surface 
topologies at oblique incidence the metasurface exhibits spatial dispersion, that is the two-
dimensional surface parameters
- in this case electric and magnetic susceptibilities and magnetoelectric parameters 
linking the in-plane electric and magnetic fields to the respective currents -
depend on the transverse momentum of the incident wave. The transverse momentum thus acts
as a self-biasing mechanism and this dependence of the surface parameters on this quantity is 
necessary for the presence of tangential (in-plane) polarization. Along
with oblique incidence, tangential
polarization was found to be necessary for the transmission to be asymmetric (dependent
 on the direction at which the metasurface is approached). As noted in Ref.~\onlinecite{tangential2},
such devices do not break time-reversal symmetry. Generally, the directionality of the 
transmission occurs whenever there is a transverse component to an incident
wave and the metasurface is spatially dispersive. The transverse momentum of
the wave incident on a spatially-dispersive metasurface functions as the 
magnetic bias in a non-reciprocal magneto-optic material. Devices may be designed
for achieving asymmetric transmission, by solving
the inverse problem, that is designing the surface parameters 
to fit the desired scattering parameters~\cite{tangential3}. Examples of non-reciprocal
phenomena which can be emulated using the transverse momentum of plane waves obliquely-incident
on a spatially-dispersive
metasurface include Faraday rotators and isolators~\cite{tangential2}. In addition, 
vortex beams carrying orbital
angular momentum carry transverse momentum even if normally incident on a metasurface and thus
devices which emulate non-reciprocal phenomena such as unidirectional transmission 
can operate even at normal incidence~\cite{tangential2}. 

Most of the studies into emulation of nonreciprocal behaviour have been conducted
on electromagnetic systems such as those described above. The few studies thus far
on acoustic systems have considered surface waves (Lamb waves) in superlattices~\cite{lamb2, lamb1}. In
Ref.~\onlinecite{lamb2} it is found that in some superlattice designs there can be a conversion between
the two modes, while the design of the superlattice can be adjusted in order to transmit
one mode of Lamb waves but not the other in a given frequency range. These two behaviours may
then be combined in order to achieve unidirectional transmission. In Ref.~\onlinecite{lamb1} the 
design's purpose is more for manipulating Lamb waves. Mode conversion is avoided
by diffracting the two modes into different propagation directions.

Bianisotropic metasurfaces~\cite{tangential2} is the name given to systems which
yield a result in one mode in response to excitation from two types of modes. For example, a magnetic
response from electric and magnetic excitation in electromagnetic systems or a transverse response
from transverse and longitudinal excitation in the case of elastic systems (``acoustic" activity~\cite{Portigal})
may be implemented by asymmetric unit cells~\cite{LiCummerNatureComm,Sieck}
 or by using $\mathcal{PT}-$symmetric systems~\cite{Zhu}. In the latter 
studies, all based
on reciprocal acoustic systems, bianisotropic metasurfaces generate different phases in the reflection
coefficient in different directions~\cite{LiCummerNatureComm}. With the addition of loss (or gain) the magnitude
of the reflection coefficient also becomes different. In all cases dealing with reciprocal materials,
the transmission coefficient in both directions is the same. Recently, elastic-wave 
propagation under oblique incidence has been studied for the transverse polarization which
is independent of the other two modes~\cite{Morini}, which are coupled under
oblique incidence and are the object of the present study. Up 
until now, studies of propagating longitudinal and transverse modes in \emph{elastic}
systems have not been made so as to achieve asymmetry in the transmission coefficients.

The addition of balanced gain (G) and loss (L) to metamaterials  constitutes
a approach to achieving reduced losses~\cite{TsironisPRL}. Parity-time ($\mathcal{PT}$), or the more general category
of pseudo-Hermitian (PH)~\cite{Deb} -symmetric
systems, despite being non-Hermitian, can have real eigenvalues, as is necessary for
propagation, as well as bound states. By modifying the system parameters, the system may
migrate from the $\mathcal{PT}$ or PH -symmetric phase into the broken phase and vice versa. 

Defects, in the form of interfacial wrinkling,
have been found to be able to control the wave propagation in layered materials by introducing complete band
gaps in 1D phononic materials~\cite{Wrinkling}. The transfer-matrix technique
has long been applied to the study of elastic-wave propagation in multilayer systems in order
to study the effect of defects or absorption~\cite{SigalasSoukoulis}. In recent work~\cite{psiachos1} we used 
transfer-matrix methods to investigate the response of a multilayered
metamaterial system containing periodicity-breaking defects to an incident acoustic plane 
wave at normal or oblique incidence. The transmission
response was found to be composed of pass-bands with oscillatory behaviour, separated by
band gaps, and covers a wide frequency range. The presence of gain and
loss in the layers was found to lead to the emergence of symmetry-breaking and re-entrant
phases.  While defects in general were found to lead to a near or complete loss of
PH symmetry at all frequencies, it was shown that they can be exploited to produce highly-sensitive responses,
making such systems good candidates for sensor applications. The presence of 
defects as well as their location within the system was
found to have a profound effect on the transmission
response: changing the
thickness of one passive layer shifts transmission resonances to different frequencies, while even
very small changes in thickness were found to produce great sensitivity in the responses. 

In this work, we extend the above study to the more general investigation of propagation of elastic waves, whereby we
exploit interconversions between longitudinal and transverse modes to optimize the directionality in not just
the reflection but also the transmission response
and also the sensitivity to periodicity-breaking defects.

\section{Methods}
We summarize our formalism from our previous work~\cite{psiachos1} making note of where it becomes
generalized owing to the consideration of both longitudinal and transverse incident modes. 

We consider a system of $n-1$ layers with normal the $\hat{z}$ direction, extending along the negative $\hat{z}$ direction
as in Fig.~\ref{fig1}. Each layer is described by homogeneous mass density $\rho$ and
elastic properties: shear modulus $\mu$ and Lam\'{e} constant $\lambda$. The
longitudinal and transverse wave speeds are thus $c=\sqrt{(\lambda+2\mu)/\rho}$ and $b=\sqrt{\mu/\rho}$ respectively.  We further
assume isotropic elasticity where the Lam\'{e} constant $\lambda=\frac{2\mu\nu}{1-2\nu}$ and the shear modulus
is $\mu=\frac{E}{2(1+\nu)}$ are expressed in terms of the Young's modulus $E$ and Poisson ratio $\nu$. A
plane wave of frequency $\omega$ is incident on the multilayer system from a general (solid or liquid) ambient medium
either normally or at an angle $\theta$ in the $xz$ plane and exits again in the same or a different medium.

The particle displacement
is in the $xz$ plane and there may be both shear and longitudinal
modes present. The
displacement field may be split into
longitudinal $\phi$ and transverse $\vec{\psi}$ potentials
\begin{equation}
\vec{u}=\vec{\nabla}\phi+\vec{\nabla}\times\vec{\psi}
\label{u}
\end{equation}
and we set $\vec{\psi}=\psi \hat{y}$. The wave equations for the potentials, assuming time-harmonic plane-waves are
\begin{align}\label{waveeq}
\nabla^2\phi+&k^2\phi=0,\;\hfill k\equiv \omega/c, \\ \nonumber
\nabla^2\psi+&\kappa^2\psi=0,\;\hfill \kappa\equiv \omega/b, 
\end{align}
and have the solutions
\begin{align}\label{prime} 
\phi_j&=\phi_j^\prime e^{i\alpha_j z}+\phi_j^{\prime\prime} e^{-i\alpha_j z},\; \hfill \alpha_j=\left(k_j^2-\xi^2\right)^{1/2},\;\hfill \xi=k_j\sin{\theta}\\ \nonumber
\psi_j&=\psi_j^\prime e^{i\beta_j z}+\psi_j^{\prime\prime} e^{-i\beta_j z},\; \hfill \beta_j=\left(\kappa_j^2-\chi^2\right)^{1/2},\;\hfill \chi=\kappa_j\sin{\theta}
\end{align}
for each layer $j$, including the terminating ambient media $j=1$ and $j=n+1,$ where the primed quantities are amplitudes. For example,
for a transverse wave incident at the half-space $n+1$, $r_l=\phi_{n+1}^{\prime}$ and $t_l=\phi_{1}^{\prime\prime}$
are the expressions for the longitudinal-mode reflection and transmission coefficients.

Through transformations from the $\{\phi^\prime,\phi^{\prime\prime},\psi^\prime,\psi^{\prime\prime}\}$ basis, a transfer matrix for 
the passage of a wave through one, and then by repeated application, through
 the whole system of $n-1$ layers may be constructed in
terms of the displacements~Eq.\ref{u}
and the stresses
\begin{align}\label{Z1}
Z_x&=\mu\left(\partial u_x/\partial z+\partial u_z/\partial x\right) \hfill \qquad \mathrm {and}\\ \nonumber
Z_z&=\lambda\left(\partial u_x/\partial x+\partial u_z/\partial z\right)+2\mu\partial u_z/\partial z
\end{align}
as
\begin{equation}
\begin{pmatrix}
u_x^{(n)}\\
u_z^{(n)}\\
Z_z^{(n)}\\
Z_x^{(n)}
\end{pmatrix}=\underline{\underline{A}}
\begin{pmatrix}
u_x^{(1)}\\
u_z^{(1)}\\
Z_z^{(1)}\\
Z_x^{(1)}
\end{pmatrix}
\label{Mmatrix2}
\end{equation}
where $\underline{\underline{A}}$ is the transfer matrix through the entire multilayer
(see ch.1, sec.8 in Ref.\onlinecite{Brekhovskikh}). 

The boundary conditions applied are the continuity of the displacements $u_x$ and $u_z,$ and that of the stresses
$Z_x,$ $Z_z$ across
the $(n,n+1)$ boundary. When this
is done, and after inverting the system of equations for the stress and displacement to obtain the amplitudes
in Eq.~\ref{prime}, we arrive at expressions for the longitudinal and transverse-mode reflections and transmissions from the incident and outgoing
sides respectively of the multilayer system.

\subsection{System}
\begin{figure}[htb]
\includegraphics[width=12cm]{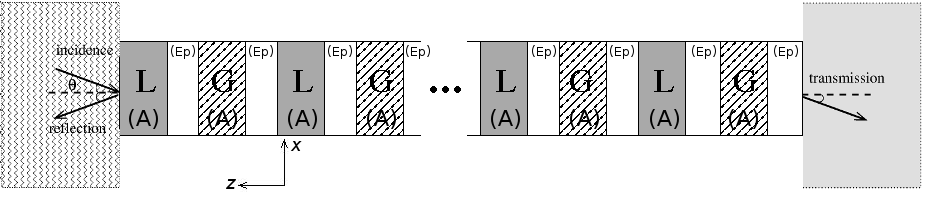}
\caption{Metamaterial system under study: oblique wave impinging onto a multilayer stack composed
of alternating gain (G) and loss (L) layers, separated by passive material. The materials 
 comprising the layers: A (alumina)
and Ep (epoxy thermoset) are discussed in the text. The angle of 
 incidence $\theta$ shown corresponds to an incident longitudinal or transverse
wave. The orientation shown corresponds
to the `Forwards' (F) direction.}
\label{fig1}
\end{figure}

In the system depicted in Fig.~\ref{fig1} we show a system of A/Ep layers from which
an elastic plane wave is incident and exits from arbitrary ambient media (half-spaces at ends). The 
propagation direction shown is the `Forwards' (F) direction. When the ambient media at 
both ends are different, or the same but not comprised of one of the materials 
of the multilayer, then the total system is spatially asymmetric, 
even without the inclusion of gain/loss (G/L) parameters.

Here, the response of multilayer system composed of alternating alumina (A) and
epoxy thermoset (Ep) layers to an obliquely-incident longitudinal or transverse wave was examined. In 
our previous study~\cite{psiachos1}, we examined the effect of incidence angle
on the response. In this study we focus almost exclusively on an incident angle of $\theta=\pi/16$ as the
results are generalizable to other angles. The parameters of the materials used 
are given in Table~\ref{parameters}. When
including G/L, through the addition of an imaginary component of positive (G) or negative (L) sign 
to the Young's modulus, we 
work in the region of balanced G/L, by imposing alternating G/L on the A layers. Loss is easily
achieved by adding dissipation but adding
gain represents a major challenge for acoustic/elastic materials. However,
piezo-electric materials have been demonstrated to be able to tune
the gain/loss (G/L) character~\cite{ChristensenPRL} of these materials. The amount of G/L imparted in our system
is in accordance with what has been demonstrated to be achievable 
for the tunable effective bulk modulus in a recent implementation of an acoustic
metamaterial~\cite{PopaPRB}.
\begin{table}
\begin{tabular}{l|| l| l| l|l|l}
Material& $E$ (GPa)& Im($E$) (GPa)& $\nu$ & $\rho$ (Mg/m$^3$)& layer thickness\\
\hline
A (alumina)&390&$\pm$20&0.26&3.9&$l$\\
Ep (epoxy thermoset) &3.5&0&0.25&1.2&$l$\\
HDPE (polyethylene)&0.7&0&0.42&0.95&-\\
Glass &65&0&0.23&2.5&-\\
Polycarbonate&2.7&0&0.42&1.2&-\\
SiC&450&0&0.15&2.8&-\\
Cermet&470&0&0.30&11.5&-\\
Cork&0.032&0&0.25&0.18&-

\end{tabular}
\caption{Parameters of the A/Ep multilayer system. $E$ (Im$E$) is the real (imaginary - when activated) part
of the Young's modulus, $\nu$ the Poisson ratio, $\rho$ the mass density, $l$ the layer thickness (uniform in this 
study). The composition of the half spaces at the ends was varied (see text) amongst Ep or one
of the other materials which follow Ep in the list.}
\label{parameters}
\end{table}

The units of the frequency $\omega$ in all the figures are $\mathrm{m}\cdot \mathrm{kHz}/l$ where $l$
is the layer thickness (see Table~\ref{parameters}) which in this study is taken to be the same for all the layers.

\subsection{Unidirectional transmission}
The transmission at some frequencies can be unidirectional. When the solid half spaces at the ends are 
identical, the forward/backward
transmissions under oblique incidence are different only a) when G/L is present, 
or b) in the presence of spatial asymmetry such as i) a periodicity-breaking
defect or ii) when the identical half spaces are composed of a material different from
those present in the multilayer (in the prototype of Fig.~\ref{fig1} studied
here). However, the transmission will be asymmetric in the two directions
only for the mode which is different from the induced mode (\textit{i.e.} not
that which impinges on the system). The asymmetric or even unidirectional transmission of the induced
mode is is a very interesting 
result which seems counterintuitive given that
the angle of transmission is technically the same as the angle of incidence. But due to the fact that 
for the induced mode there is no incident
wave, the outgoing angle for the induced mode is in general different than that of the 
incident mode. In this study in addition to asymmetry, we also examine the effect of
periodicity-breaking defects. The defect need not be liquid for the asymmetry in the transmission
response to occur, although in this
study we consider only liquid periodicity-breaking defects. 

The surface band structure
of the system depicted in Fig.~\ref{fig1} is of the same type as those depicted in Fig.2 of 
Ref.~\onlinecite{psiachos1} but with the horizontal axis transformed to show $k_x,$ the
wavevector in the plane of the layers, equal to $k\sin{\theta}$
 where the wavevector $k$ is for propagation in both the $x$ and $z$ directions and
corresponds to the solutions of the dispersion relation analogous to those shown in Fig.1 of 
Ref.~\onlinecite{psiachos1}. It is clear that for the cases of oblique incidence $\theta\neq 0$ 
such a transformation results in 
spatial dispersion, a precondition for unidirectional transmission~\cite{tangential1}.

\section{Results}
In this section we present our main results upon changing either the composition of the half-spaces 
or adding defects. The purpose is to highlight the directionality
in the transmission and reflection responses. When the half spaces admit only
longitudinal waves, \textit{i.e.} for liquid ambient media (see for example
the systems investigated in 
Ref.~\onlinecite{psiachos1}, no asymmetry in the transmission is found, while
the magnitude of the reflections is different only upon the inclusion of G/L elements
(PH symmetric system)~\cite{psiachos1}. Both longitudinal and transverse modes 
need to be transmitted for there to be an asymmetry in the transmission responses. When the
half-spaces admit both longitudinal and transverse modes, \textit{i.e.} are solid, then directionality 
in the reflection is seen when the total system is asymmetric in space (\textit{viz.} Fig.~\ref{fig1}
when the ambient media are different or not the same as either of the materials in the multilayer). However,
the response to the same mode type as the incident wave showed no directionality when the half spaces were 
identical. When the half spaces were different, the response to either the incident or the induced mode 
was found to display a directionality as the incident and outgoing
angle for this type of mode were no longer identical.

\begin{figure}[!htb]
\includegraphics[width=10cm]{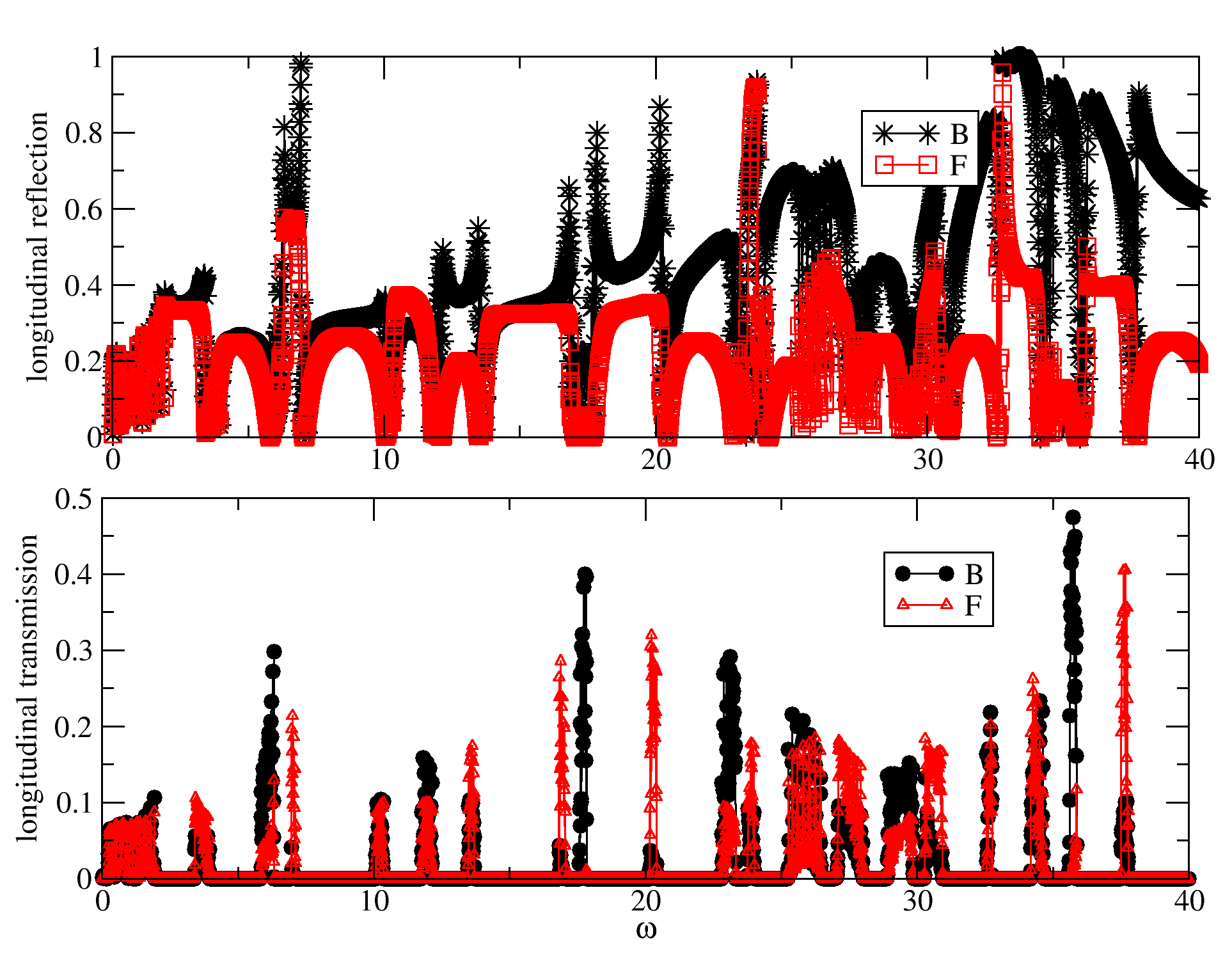}
\caption{Longitudinal-mode response for a A/Ep multilayer system (16 total layers) with no
G/L present between two glass 
half-spaces on which a transverse
mode is incident at an angle of $\pi/16$ in either
the forward (F) or backwards (B) direction.}
\label{fig2}
\end{figure}

\begin{figure}[!htb]
\includegraphics[width=10cm]{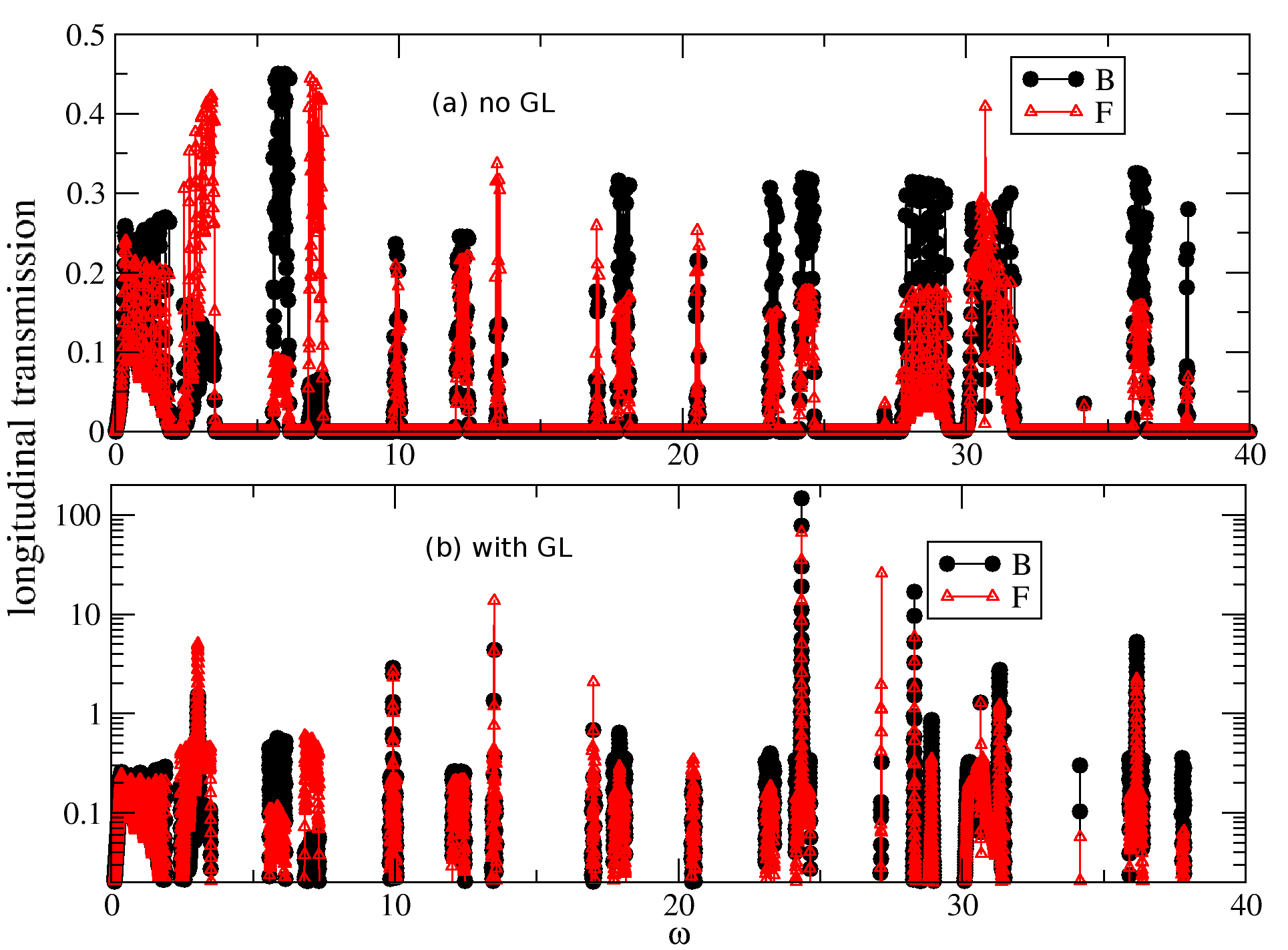}
\caption{Longitudinal-mode response for an A/Ep multilayer system (16 total layers) between
two glass half-spaces on which
a transverse mode is incident at an angle of $\pi/8$ in either the forwards (F) or backwards (B)
direction, with (a) no
G/L present and (b) with GL.}
\label{fig3}
\end{figure}

 In Fig.~\ref{fig2} we show the reflection and transmission when both half spaces
are glass and without any G/L present for a transverse mode incident 
at an angle of $\pi/16$. The reflections are different because
the system is asymmetric and the ends are solid enabling the interconversion between modes for 
the incoming and outgoing waves -\textit{viz.} asymmetric system
but liquid ambient media in Ref.~\onlinecite{psiachos1} where the reflections showed directionality
only when G/L was present. There are several narrow frequency ranges in which the transmission in 
particular is virtually unidirectional, namely around $\omega=6.1, 7.0, 16.9, 17.7, 20.3, 35.7, 
37.6$. Similarly for Fig.~\ref{fig3}a where the
angle is now $\pi/8$. When G/L is added however, as in Fig.~\ref{fig3}b, many of these resonances
are not responsive and remain at the same low level. In particular those narrow bands centred 
around $\omega=$ 5.8 and 7.1, displaying strong unidirectional transmission did not respond 
to G/L. The responses at other frequencies are
 amplified in both directions rather than being attenuated in one direction (\textit{e.g} at 
$\omega=$ 3.0, 36.2) so these are obviously not 
useful for operation as a unidirectional device. In the case of 
Fig.~\ref{fig3}b the system, extrapolated to an infinite multilayer, 
is in the broken PH regime for all frequencies as the angle is relatively large.

\begin{figure}[!htb]
\includegraphics[width=10cm]{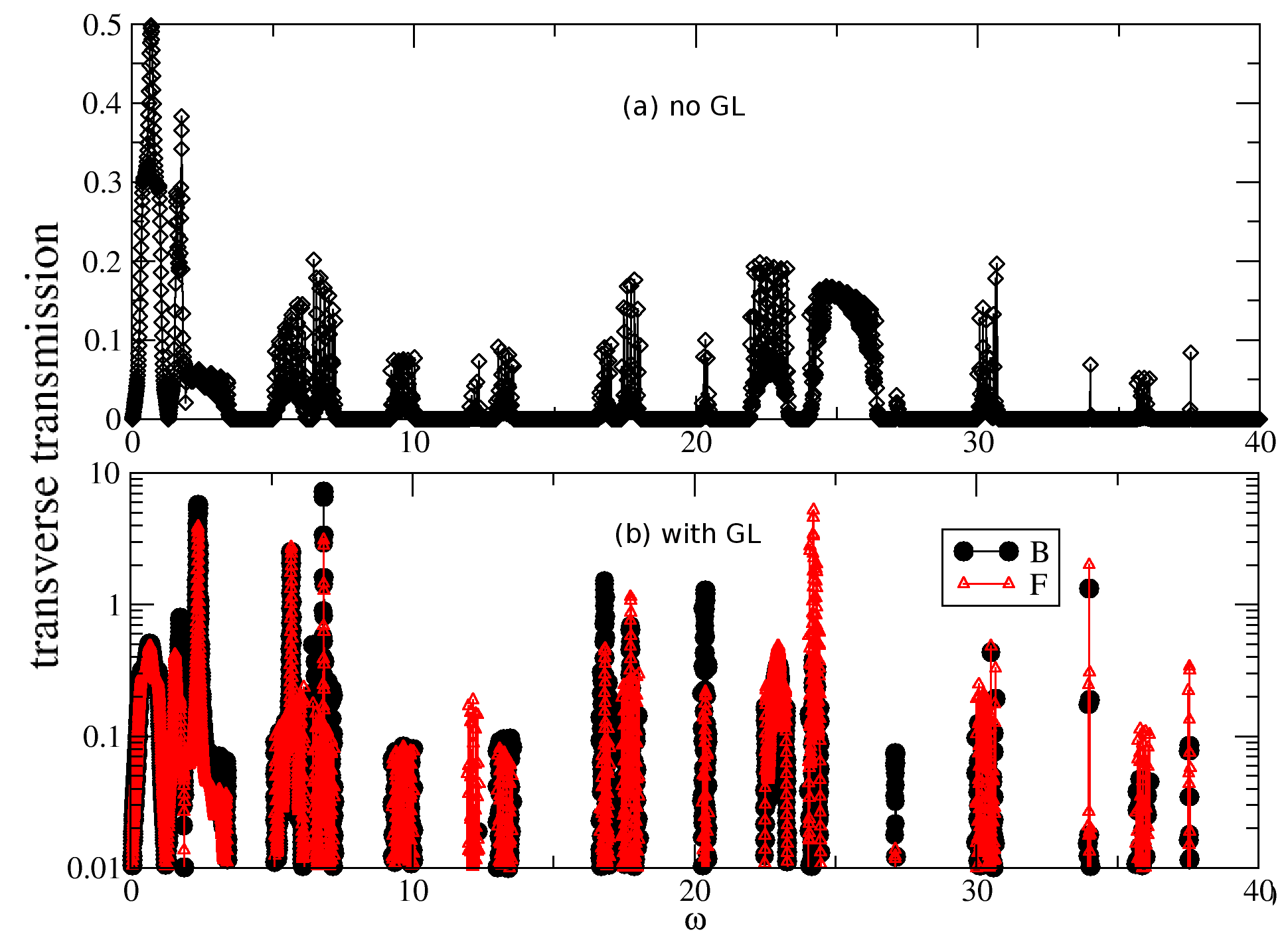}
\caption{Transverse-mode transmission response for a A/Ep multilayer system (16 total layers): a) without G/L
and b) with G/L, between two Ep half-spaces on which a longitudinal
mode is incident at an angle of $\pi/16$ in either
the forward (F) or backwards (B) direction. In a) both directions give the same result.}
\label{fig4}
\end{figure}

In Fig.~\ref{fig4} we show the transverse-mode 
transmission response of the same A/Ep system located between Ep at both ends
to an incident longitudinal wave. Since the total system, without the inclusion 
of G/L (case of Fig.~\ref{fig4}a) here is symmetric, the transmission is 
independent of direction. Once G/L is added however (Fig.~\ref{fig4}b), notable asymmetry in the
transmission at some frequencies occurs. In particular, at $\omega=$1.76, the F direction gives
a transmission of 0.18 while the B direction gives 0.82 but this behaviour is not seen over
a wide range, as it is caused by a shift in overlap in the transmission spectra of the two
directions. Other cases of strongly-asymmetric
transmission, such as the band around $\omega=$12, show low absolute values overall, in this case
a value of 0.2 for the F direction and 0.01 for the B direction, making them
not very exploitable. In the band around $\omega$=24.2, the transmission of the F direction reaches 5.3
and that of the B direction 0.36 in this example, making the transmission not quite unidirectional 
but highly asymmetric. In the case of Fig.~\ref{fig4}b there are some regions of 
$\omega$ where an analysis of the transfer matrix shows that the system is in the
PH-symmetric regime.

\begin{figure}[!htb]
\includegraphics[width=10cm]{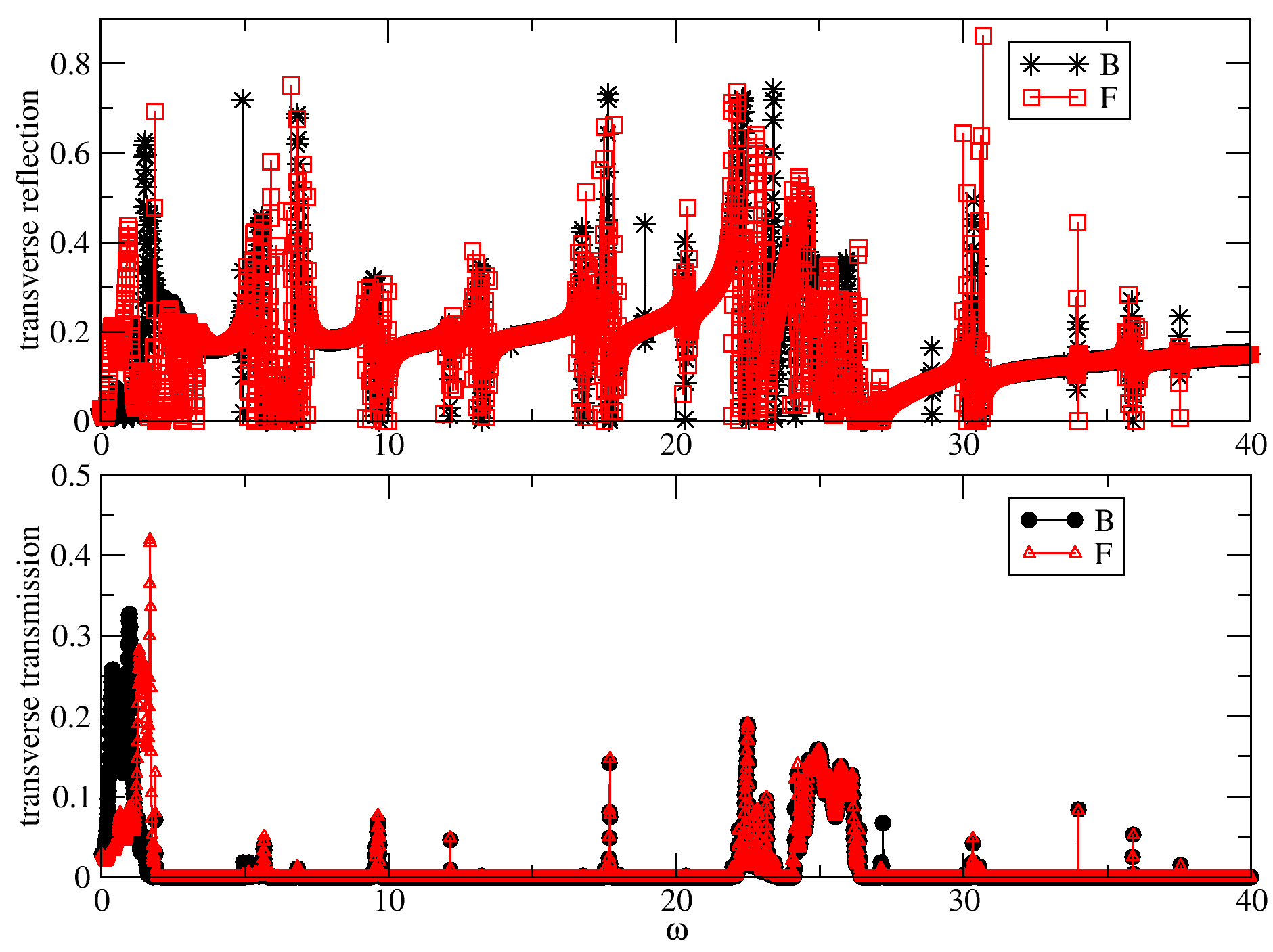}
\caption{Same system and incident conditions as Fig.~\ref{fig4}a but with water replacing the second Ep, as taken from 
the F direction. Shown are both the transverse-mode reflection and transmission responses.}
\label{fig5}
\end{figure}

\begin{figure}[!htb]
\includegraphics[width=10cm]{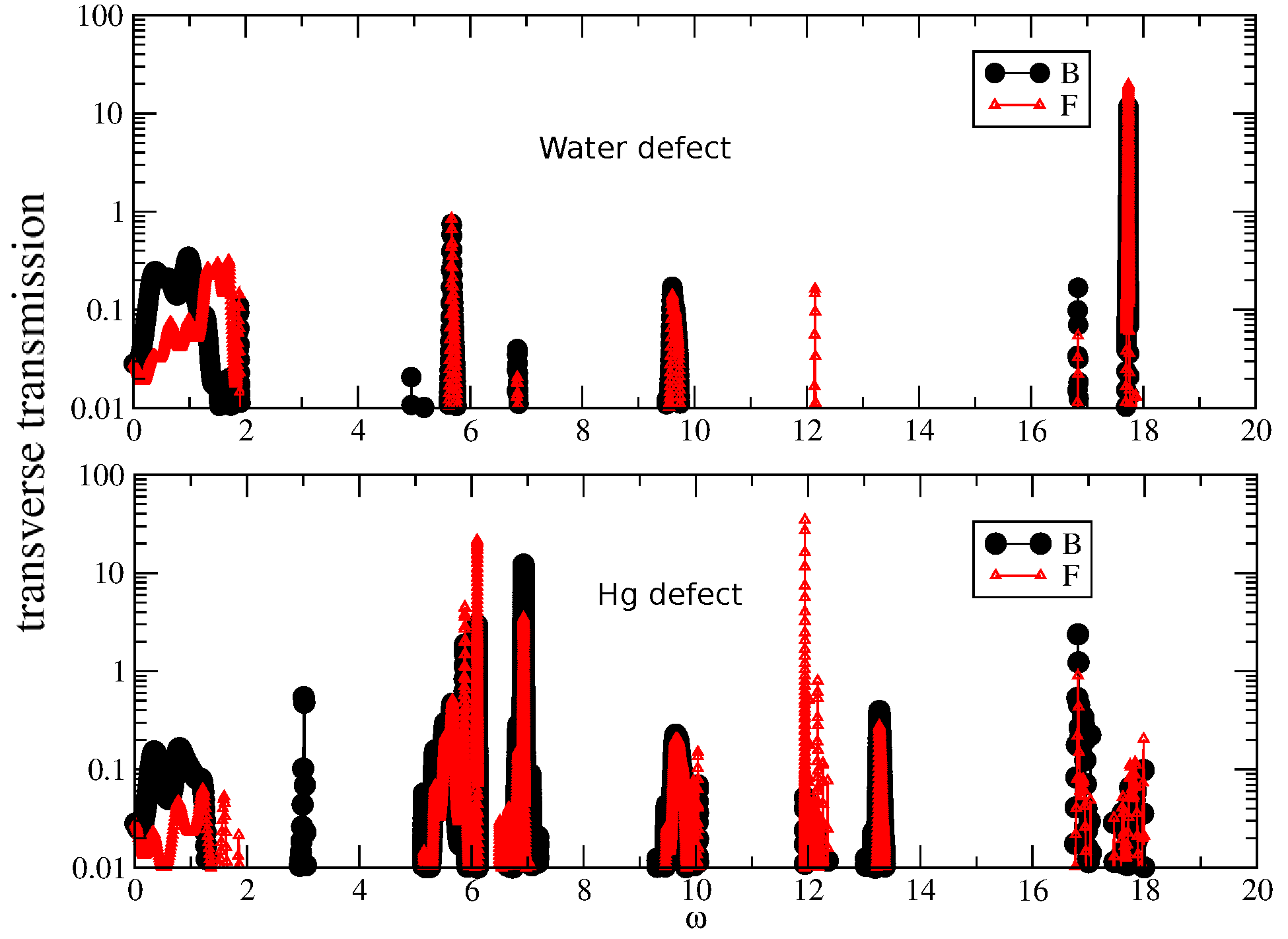}
\caption{Transverse-mode transmission response for the same system and conditions as Fig.~\ref{fig4}b but 
with water or Hg replacing the second Ep, as taken from 
the F direction.}
\label{fig6}
\end{figure}

In Fig.~\ref{fig5} we show the transverse transmission and reflection responses of the otherwise passive system, 
but where we have also added a water periodicity-breaking defect replacing the second Ep in the F direction. The 
incidence is again of a longitudinal mode, at an angle of $\pi/16.$ At low $\omega,$
clear directionality is evident especially in the transmission, demonstrating that even if the
total system including the ambient media is symmetric, the addition of a periodicity-breaking defect
can generate asymmetry in the transmission and the reflection responses. However, the direction
with the high transmission has the low reflection and vice versa, rather than
the system being unidirectional in both transmission and reflection responses. When 
a defect of water or Hg is added
to the system Fig.~\ref{fig4}b - including G/L - we can achieve large responses 
as in Fig.~\ref{fig6}. However, the responses here, particularly for the case of water,
were found to be highly variable with respect
to defect location, successively alternating between large responses or 
sinking to below 1 depending on which Ep was replaced. What is most notable is that the response
with the Hg defect is not only greatly magnified but 
highly unidirectional at $\omega$=12, with one direction having a transmission 
of 0.05 and the other at 33. This same peak is also present in the undefected system 
(Fig.~\ref{fig4}b) where it is asymmetric but nevertheless has a weak response. With the water defect
this peak maintains a similarly-low response while in one direction it is nonexistent on the scale used. The cases with defects studied (Fig.~\ref{fig6}) were in the broken PH phase at all frequencies.

\begin{figure}[htb]
\includegraphics[width=12cm]{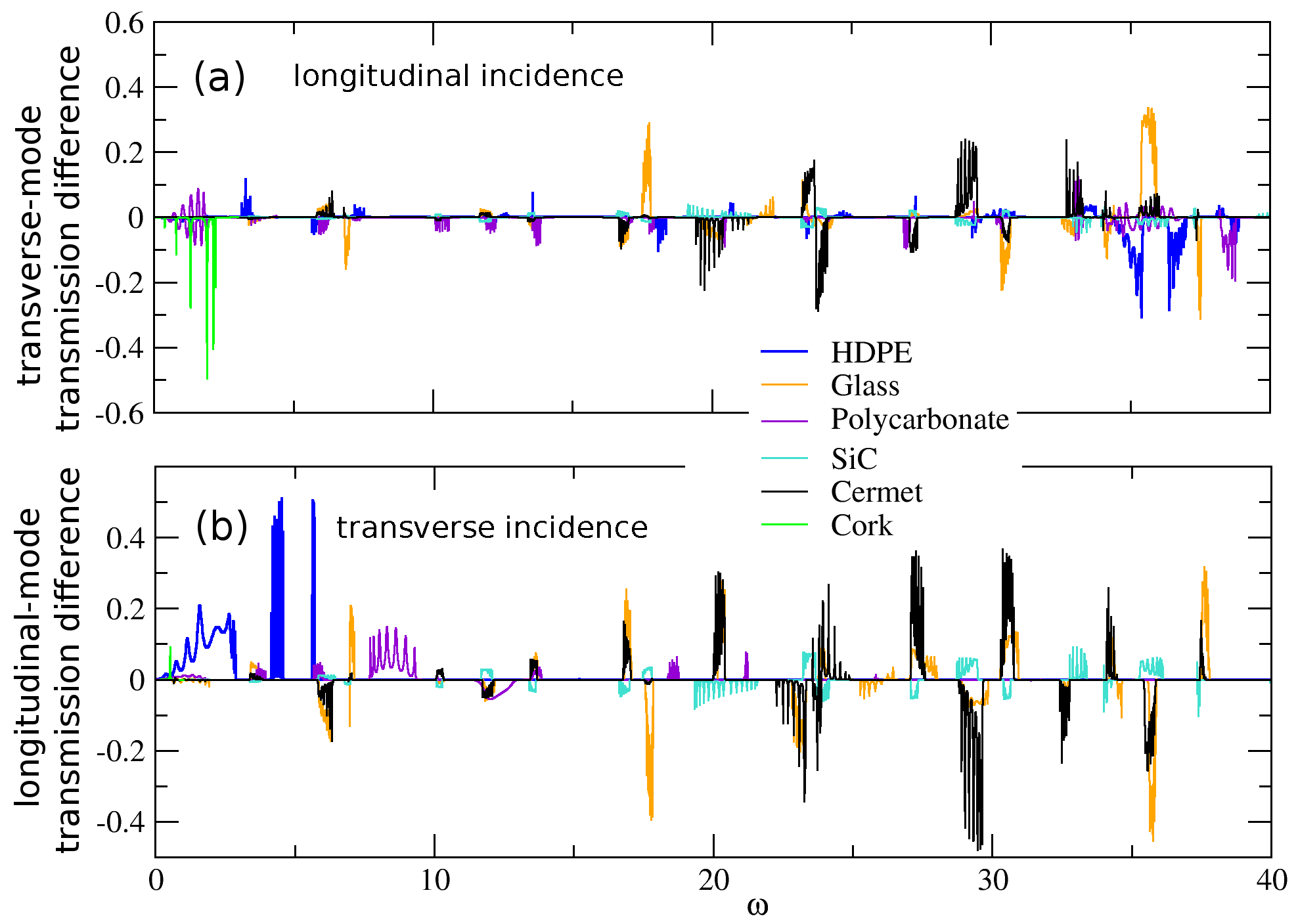}
\caption{Transverse-mode (a) and Longitudinal-mode (b) transmission response difference
between F and B directions for a A/Ep multilayer system 
(16 total layers) without G/L, between two half-spaces of varying composition. In (a) a longitudinal
mode is incident at an angle of $\pi/16$ while in (b) a transverse mode is incident at $\pi/16.$}
\label{fig7}
\end{figure}

\begin{figure}[htb]
\includegraphics[width=12cm]{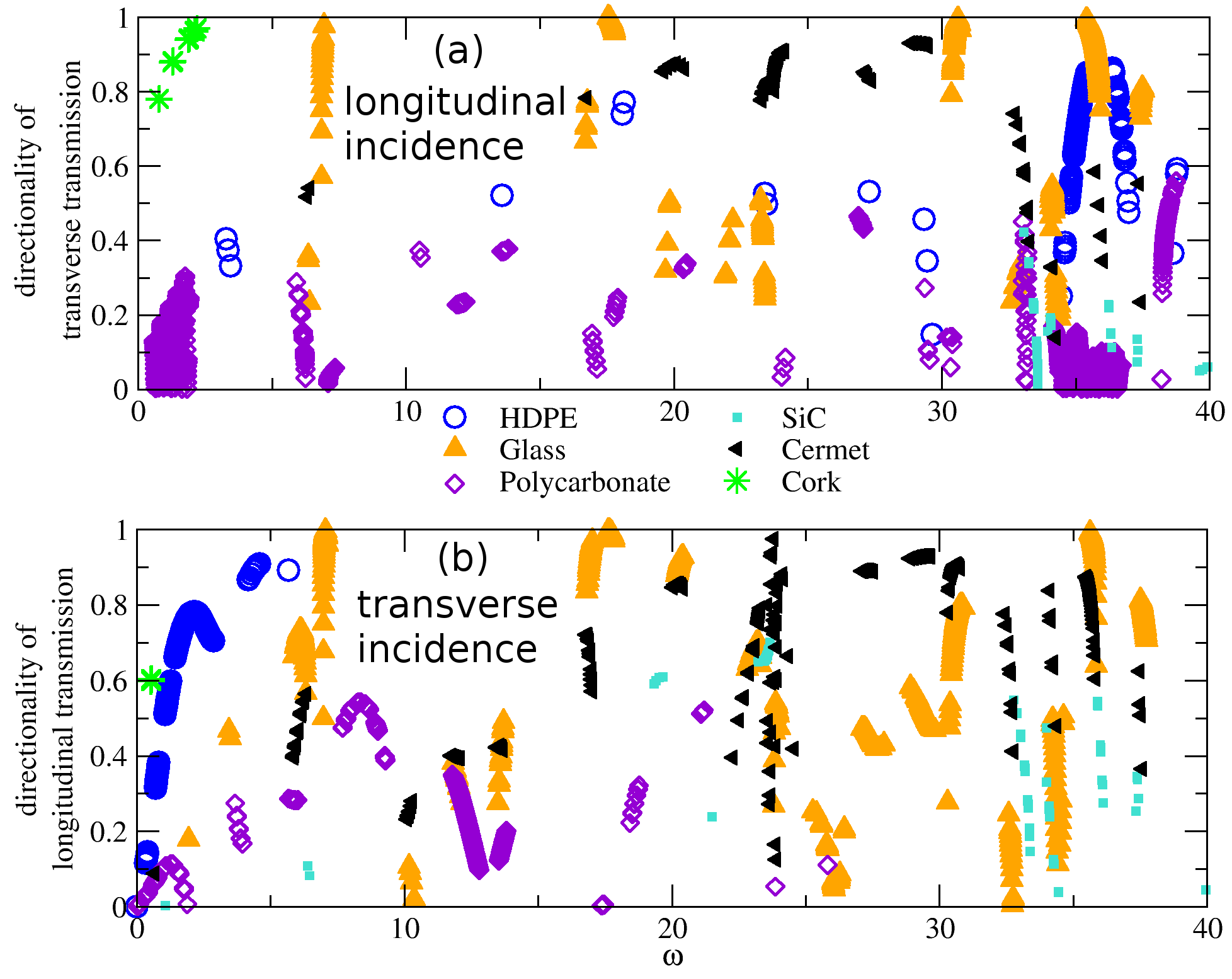}
\caption{Transverse-mode (a) and Longitudinal-mode (b) transmission response relative directionality
	between F and B directions for the same system depicted in Fig.~\ref{fig7}}
\label{fig8}
\end{figure}

In Fig.~\ref{fig7} we depict the extent of the directional response of the induced-mode response to
an incident longitudinal or transverse wave at an angle of $\pi/16$ for the periodic
multilayer system without any defects, when the ambient media, although the same
at both ends, are varied. More specifically, in this figure we show the 
difference in the transmission response of the induced mode between 
the forward and backward directions. We see that for longitudinal
incidence, cork ends give the largest asymmetry in the response, and SiC the smallest. For 
transverse incidence, cork
gives almost no difference between the two directions while HDPE gives the largest 
difference. At $\omega=6$ and 20 in Fig.~\ref{fig7}b the peak
for glass is barely visible, behind that of cermet but at the same 
depth. Similar 
plots but for relative directionality, \textit{i.e.} the difference in the 
transmissions with respect to the largest of the two transmissions, may 
lead to misleading conclusions wherein for frequencies where both transmissions are low
but one is essentially zero the result yields perfect directionality while situations 
with a higher absolute difference, in which one transmission is high and the 
other is merely low, a lower value of the directionality results. We plot
the relative directionality in Fig.~\ref{fig8}, for the same systems as in Fig.~\ref{fig7}. We
have included only data where one of the two transmission values exceeds a `cutoff', set
here as 0.1, in order
to avoid fictitiously high values of directionality for the case mentioned above, where
both transmissions are very low. In the plots of Fig.~\ref{fig8}, glass in particular
yields several perfect directionalities at the same frequencies where some
small differences between the two directions are seen, but this is 
due to the the fact
that one of the directions had a near zero response while the other had a value
which was low but above the cutoff used.

\begin{figure}[htb]
\includegraphics[width=12cm]{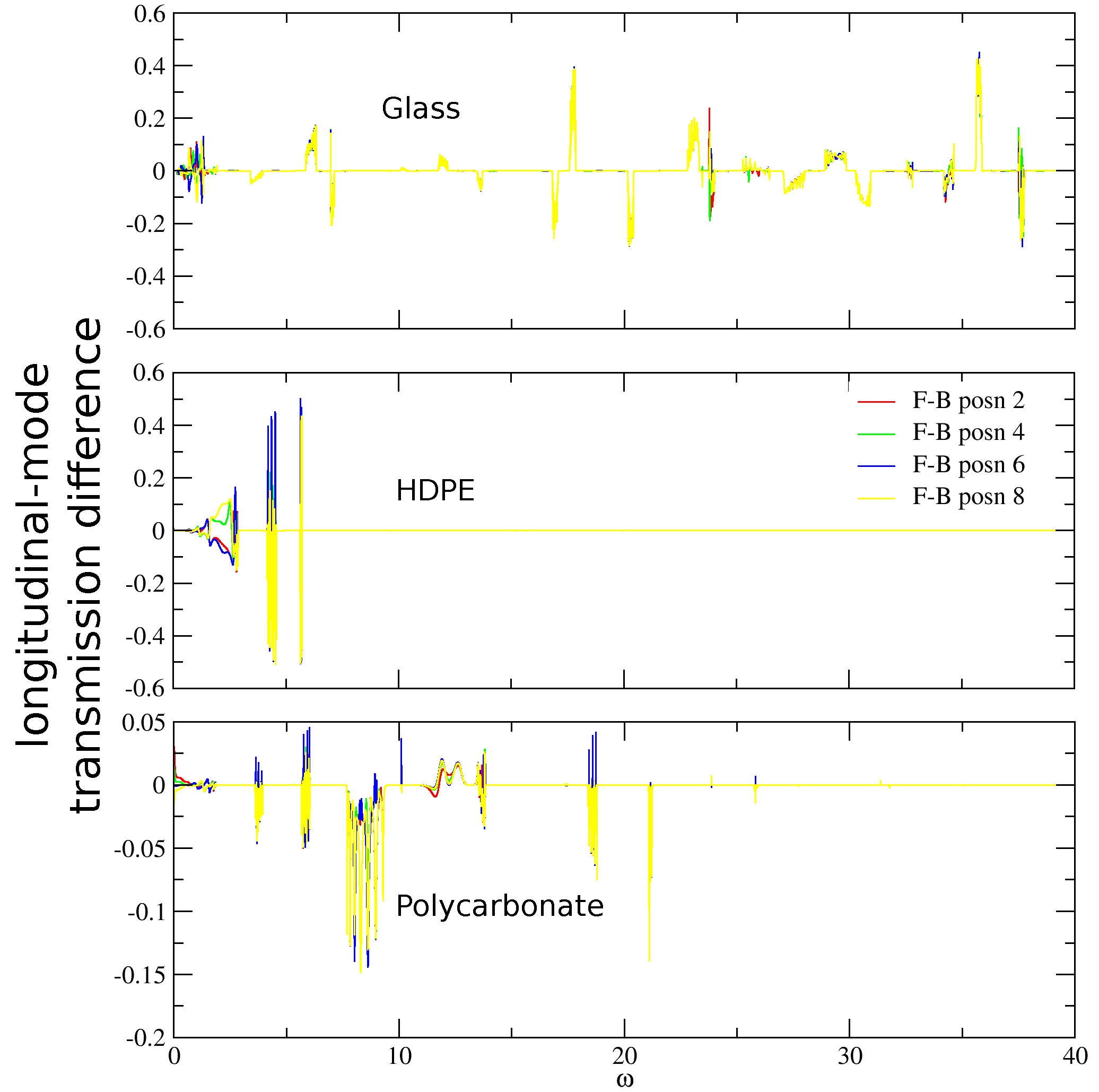}
\caption{Longitudinal-mode transmission response difference
between F and B directions and compared to the respective undefected system for a transverse mode 
incident at an angle of $\pi/16$ onto a A/Ep multilayer system 
(16 total layers) without G/L, between two half-spaces of glass, HDPE or polycarbonate.}
\label{fig9}
\end{figure}

In Fig.~\ref{fig9} we show the difference in the defect response between
the two directions for three materials: 
glass, HDPE and polycarbonate, to an incident transverse wave at $\pi/16,$ that is how does
the presence of a defect, in this case water replacing Ep in different locations, 
affect the asymmetry in the response in the two directions. Having an 
Hg defect rather than a water defect did not significantly
alter the results, as Hg shows an improvement in defect response compared with water 
but not in the directional asymmetry of the response, something which is more affected by the material
at the ends. These plots were produced by taking the difference between the defect responses
(defined as the difference between the response of the system containing a defect and that without)
of the F and the B directions. HDPE showed
the highest response and polycarbonate almost none meaning that the difference
in the responses between the two directions due to the presence of a defect
is mainly determined by the composition of the ends, as in the case of
no defects (\textit{viz.} Fig.~\ref{fig7}b for transverse incidence). It is interesting however, that sometimes there is a great
variation even in the sign of the response depending on the location of the defect.

\begin{figure}[htb]
\includegraphics[width=10cm]{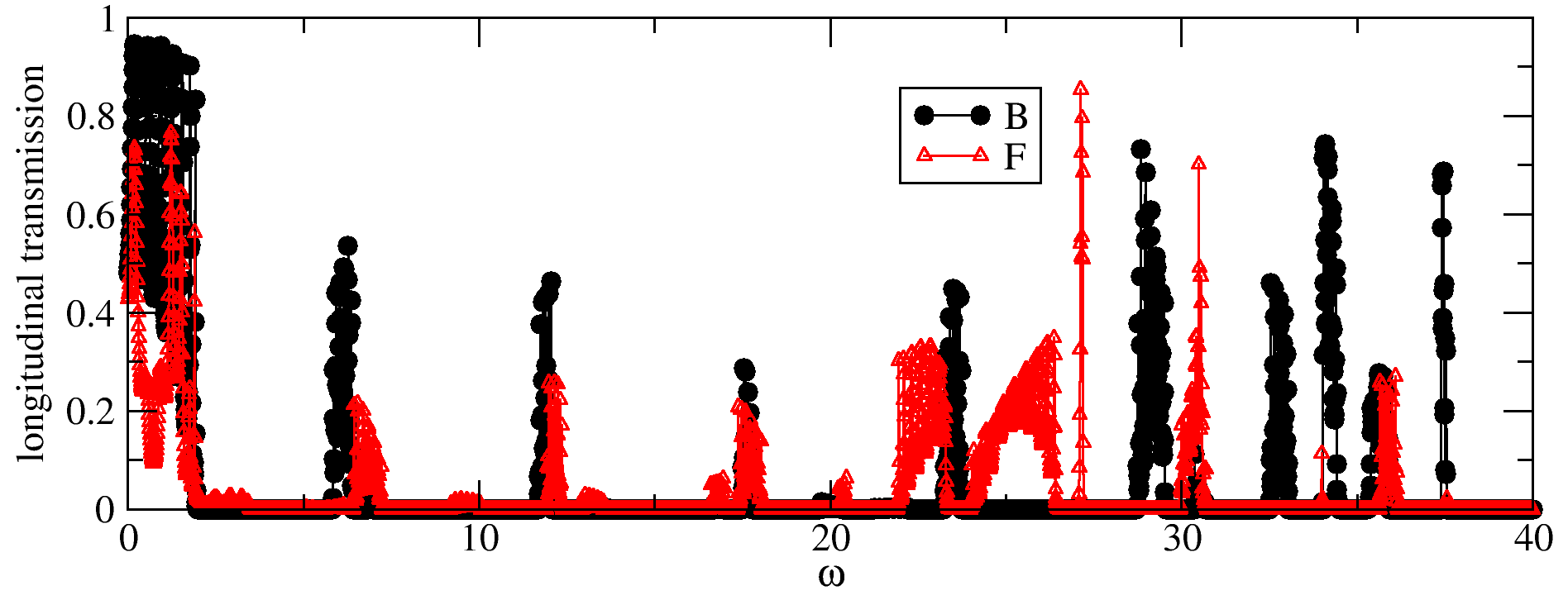}
\caption{Longitudinal-mode transmission response for an A/Ep multilayer system
(16 total layers) without G/L, between Ep/Glass half-spaces (referring to the F direction) on which a longitudinal
mode is incident at an angle of $\pi/16$ in either
the forwards (F) or backwards (B) direction.}
\label{fig10}
\end{figure}

\begin{figure}[htb]
\includegraphics[width=10cm]{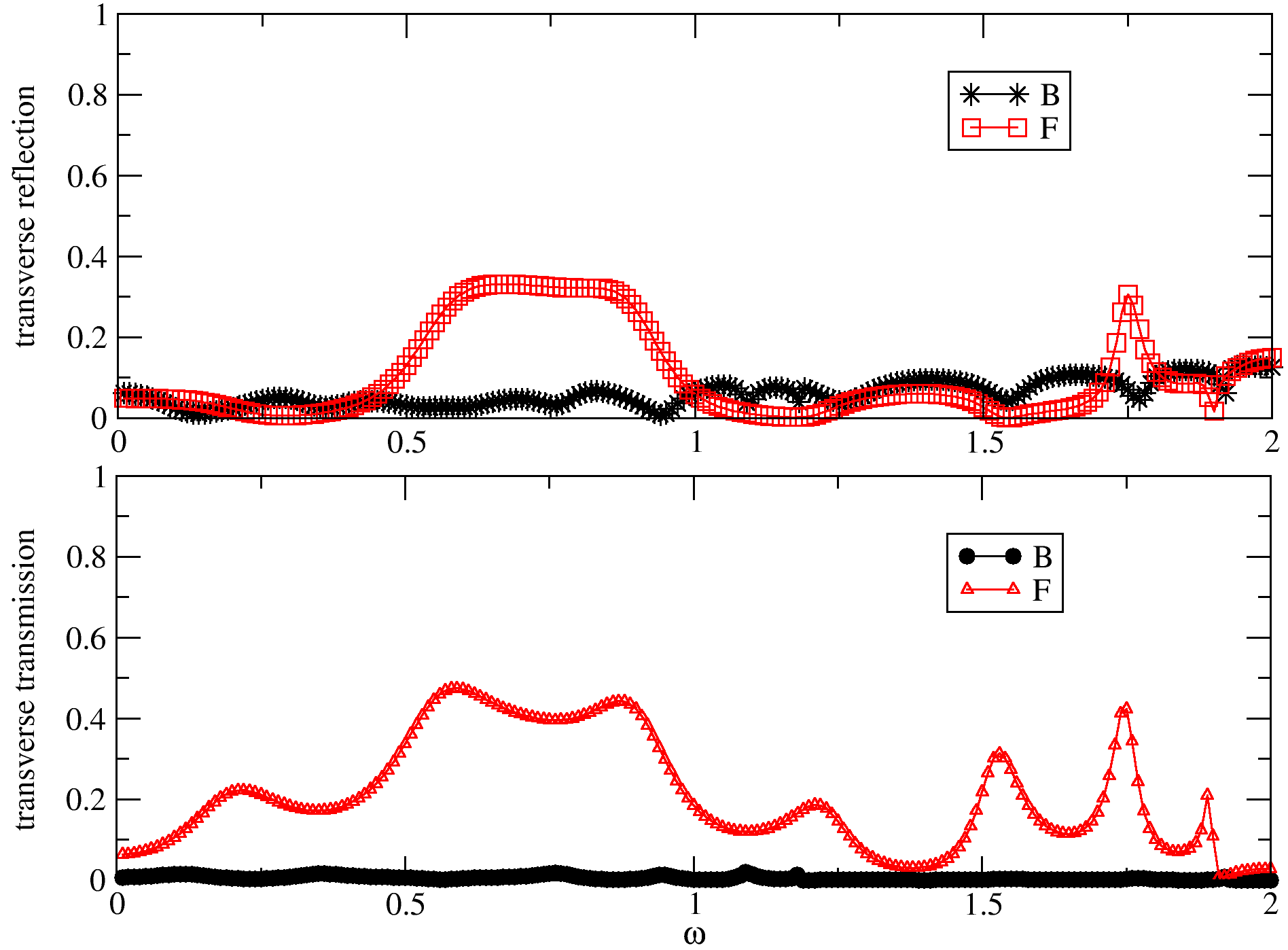}
\caption{Transverse-mode reflection and transmission responses for a A/Ep multilayer system
(16 total layers) without G/L,
between Ep/Glass half-spaces (referring to the F direction) on which a longitudinal
mode is incident at an angle of $\pi/16$ in either
the forwards (F) or backwards (B) direction.}
\label{fig11}
\end{figure}

We also examined the effects of having different half spaces at the ends. In all of the previous cases
where the ambient media were the same, the transmission response of the same mode as the incident was unaffected by direction. However, when the ambient media are different, the transmission response displays
a directional asymmetry for this mode. As an example, we show in Fig.~\ref{fig10} the longitudinal-mode
transmission response to a longitudinal incident mode at $\theta=\pi/16$ for same multilayer system as before
without any G/L or defects. In general, we find that when the half-spaces are different 
then the asymmetry in the response between the two directions is much more evident 
owing to the fact that for example in this case
 the transmission angle is different
from the incident angle. In Fig.~\ref{fig11} we show, for the same system, the transverse-mode response to the longitudinally incident wave and here we 
find a region with asymmetry simultaneously in both 
the transmission and reflection responses.

Lastly, we briefly mention that if just one of the half-spaces is solid and the other liquid,
we still have asymmetry in the longitudinal-mode reflection and transmission responses, without any G/L or defects
present, in the case of an incident longitudinal wave. The transverse-mode response to
a longitudinal incident mode is a trivial case, resulting in a total suppression of the transmission response 
in one direction. This situation is more likely
to be encountered in an application, for example wave-breaking, while the case of both ends being solid
is more applicable to seismology.

\section{Conclusions}
We have investigated the propagation of obliquely-incident longitudinal and
transverse elastic waves in a completely passive multilayered
metamaterial system having spatial dispersion, located 
between solid ambient media. We find that 
the interconversion between
longitudinal and transverse modes within the system leads to large responses in 
the induced (different from the incident) mode. We further find that 
there can be directionality in not just the reflection response but 
also in the transmission for the induced mode. This represents
an emulation of a property of non-reciprocal propagation and it is achievable for a completely
passive metamaterial system if both transverse and longitudinal modes are able to propagate,
as is the case for when the elastic waves originate from and end up in solid ambient media.
 
In some cases, and
especially when the elastic wave enters and exits from different solid materials,
we find that the transmission and reflection responses can be simultaneously unidirectional, with
both the transmission and reflection responses being strongly suppressed in one direction and 
greatly enhanced in the other. If spatial asymmetry in the total system including the ambient
media is present, then the asymmetry in the transmission occurs
even in the absence of periodicity-breaking defects or gain and loss in the layers although
it can be greatly modified by including these parameters. The 
degree of asymmetry in the transmission response, whether or not a periodicity-breaking defect is present, is determined 
mainly by the composition of the ambient media from which the elastic wave enters or exits the multilayer system. However, 
there was great variability in the response depending on the location of the defect within the multilayer 
system as well as some variation in the absolute sensitivity with the type of defect present. Devices 
built on this property could for example, function as sensors. The 
introduction of gain or loss in the system causes some of the responses to be greatly enhanced and sometimes
profoundly magnifies their asymmetry. Some uses for the unidirectional transmission and reflection properties
displayed by the system studied here could include for example, breakwater structures and seismic protection.

\end{document}